\documentclass[%
 reprint,
 amsmath,amssymb,
 aps,
]{revtex4-1}

\usepackage{graphicx}% Include figure files
\usepackage{dcolumn}% Align table columns on decimal point
\usepackage{bm}% bold math
\pdfoutput=1
\begin{document}

\preprint{APS/123-QED}

\title{Composite Fermions Waltz to the Tune of a Wigner Crystal}% Force line breaks with \\

\author{H.\ Deng, Y.\ Liu, I.\ Jo, L.N.\ Pfeiffer, K.W.\ West, K.W.\ Baldwin, and M.\ Shayegan}

\affiliation{%
  Department of Electrical Engineering, Princeton University
}

\date{\today}% It is always \today, today,

\begin{abstract}
When the kinetic energy of a collection of interacting two-dimensional (2D) electrons is quenched at very high magnetic fields so that the Coulomb repulsion dominates,
the electrons are expected to condense into an ordered array, forming a quantum Wigner crystal (WC).
Although this exotic state has long been suspected in high-mobility 2D electron systems at very low Landau level fillings ($\nu<<1$),
its direct observation has been elusive. Here we present a new technique and experimental results directly probing the magnetic-field-induced WC.
We measure the magneto-resistance of a bilayer electron system where one layer has a very low density
and is in the WC regime ($\nu<<1$), while the other (''probe'') layer is near $\nu=1/2$ and hosts a sea of composite fermions (CFs).
The data exhibit commensurability oscillations in the magneto-resistance of the CF layer, induced by the periodic potential of WC electrons in
the other layer, and provide a unique, direct glimpse at the symmetry of the WC, its lattice constant, and melting.
They also demonstrate a striking example of how one can probe an exotic many-body state of 2D electrons
using equally exotic quasi-particles of another many-body state.
%\begin{description}
%\end{description}
\end{abstract}

\pacs{Valid PACS appear here}% PACS, the Physics and Astronomy
\maketitle

%\tableofcontents

Interacting 2D electrons subjected to high perpendicular magnetic fields ($B$) and cooled to low temperatures
exhibit a plethora of exotic quasi-particles and states \cite{Tsui.PRL.1982, Note1, Note2, Jain.CF.2007}.
At $\nu=1/2$ Landau level filling factor, for example,
the interacting electrons capture two flux-quanta each
and create new quasi-particles, the so-called CFs \cite{Jain.CF.2007, Jain.PRL.1989, Halperin.PRB.1993},
which behave as essentially non-interacting particles.
The CFs offer an elegant explanation for the fractional quantum Hall effect (FQHE)\cite{Tsui.PRL.1982, Note1, Note2, Jain.CF.2007, Jain.PRL.1989, Halperin.PRB.1993}.
Furthermore, because of the flux attachment,
the effective magnetic field ($B_{eff}$) felt by the CFs
vanishes at $\nu=1/2$ so that CFs occupy a Fermi sea and exhibit Fermi-liquid-like properties,
similar to their zero-field electron counterparts
\cite{Jain.CF.2007, Halperin.PRB.1993, Willett.PRL.1993, Kang.PRL.1993, Smet.PRB.1997, Smet.PRL.1998, Kamburov.PRL.2014}.
In particular, in the presence of an imposed periodic potential,
CFs' resistance oscillates as a function of $B_{eff} = B-B_{1/2}$,
where $B_{1/2}$ is the value of the external field at $\nu=1/2$
\cite{Kang.PRL.1993, Smet.PRB.1997, Smet.PRL.1998, Kamburov.PRL.2014}.
These oscillations are a signature of the commensurability of CFs' quasi-classical cyclotron orbit diameter
with an integer multiple of the period of the imposed potential,
and their positions in $B_{eff}$ reflect the symmetry and period of the potential.

Another example of a collective state is the WC \cite{Wigner.PR.1934},
an ordered array of electrons, believed to form at very small fillings ($\nu << 1$)
when the Coulomb repulsion between electrons dominates
\cite{Note3, Lozovik.JETP.1975, Lam.PRB.1984, Levesque.PRB.1984, Note3, Andrei.PRL.1988, Jiang.PRL.1990, Goldman.PRL.1990, Li.PRL.1991, Li.PRL.1997, Chen.NatPhys.2006, Tiemann.NatPhys.2014}.
The WC, being pinned by the ubiquitous residual disorder,
manifests as an insulating phase in DC transport \cite{Note3, Jiang.PRL.1990, Goldman.PRL.1990, Li.PRL.1991},
and exhibits resonances in its AC (microwave) transport which strongly suggest collective motions of the electrons
\cite{Note3, Andrei.PRL.1988, Li.PRL.1997, Chen.NatPhys.2006}.
So far, however, there have been no direct measurements of the WC order or its lattice constant.

\begin{figure}[hbp]
\includegraphics[width = 0.45\textwidth]{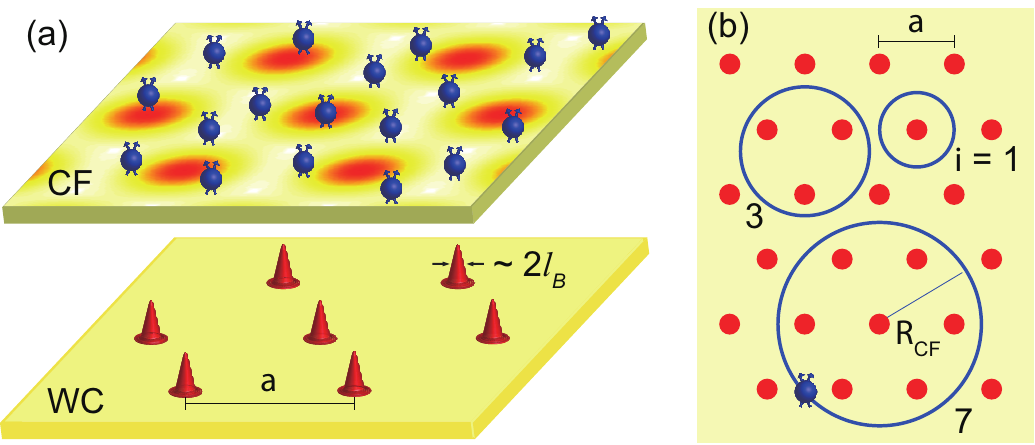}
\caption{
Overview of our measurements.
(a) Our bilayer system has a high-density top-layer which hosts a CF Fermi sea
at high magnetic fields near its filling factor $\nu_T=1/2$.
The bottom-layer has a much lower density so that it is at a very small filling factor ($\nu_B << 1$),
thus allowing a WC to form. The top-layer feels a periodic potential modulation
from the bottom WC layer's charge density; $l_B$ is the magnetic length.
(b) As the magnetic field is swept near top-layer's 1/2 filling,
the CFs in the top-layer execute cyclotron motion,
leading to commensurability maxima in the magneto-resistance
of the top-layer when the CF cyclotron orbit encircles $1, 3, 7, \cdots$ lattice points.
}
\end{figure}

Here we present high magnetic field data in a bilayer electron system with unequal layer densities.
One layer has a very low density and is in the WC regime,
while the adjacent layer is near $\nu=1/2$ and contains CFs (Fig. 1).
The CFs feel the periodic electric potential of the WC in the other layer and exhibit magneto-resistance maxima
whenever their cyclotron orbit encircles certain integer number of the WC lattice points.
The positions of the maxima are consistent with a triangular WC and yield a direct measure of its lattice constant,
while the disappearance of the resistance maxima as the temperature is raised signals the melting of the WC.

Our samples were grown via molecular beam epitaxy and consist of two,
30-nm-wide, GaAs quantum wells (QWs) separated by a 10-nm-thick,
undoped Al$_{0.24}$Ga$_{0.76}$As barrier layer.
The QWs are modulation-doped with Si $\delta$-layers asymmetrically:
the bottom and top spacer layer thicknesses are 300 nm and 80 nm respectively.
This asymmetry leads to very different as-grown 2D electron densities in the QWs;
the top-layer has a density of $\approx$ 1.5 $\times 10^{11}$ cm$^{-2}$,
much higher than the bottom-layer density $\approx 0.3 \times 10^{11}$ cm$^{-2}$.
Based on the growth parameters and our data for other, single-layer samples,
we estimate the top and bottom layers to have low-temperature mobilities of $\approx 10^{7}$
and $\approx10^{6}$ cm$^{2}$/Vs, respectively.
Here we report data for a Hall bar sample (200 $\mu$m wide and 800 $\mu$m long) with InSn ohmic contacts.
The sample was thinned down to about 120 $\mu$m
and then fitted with an In back-gate which covers its entire back surface.
Applying a negative voltage bias ($V_{BG}$) to the back-gate reduces the bottom-layer density.
The measurements were carried out in a dilution refrigerator
with a base temperature of $\approx  30$ mK,
and using low-frequency ($\leq$ 40 Hz) lock-in technique.

\begin{figure*}[htbp]
\includegraphics{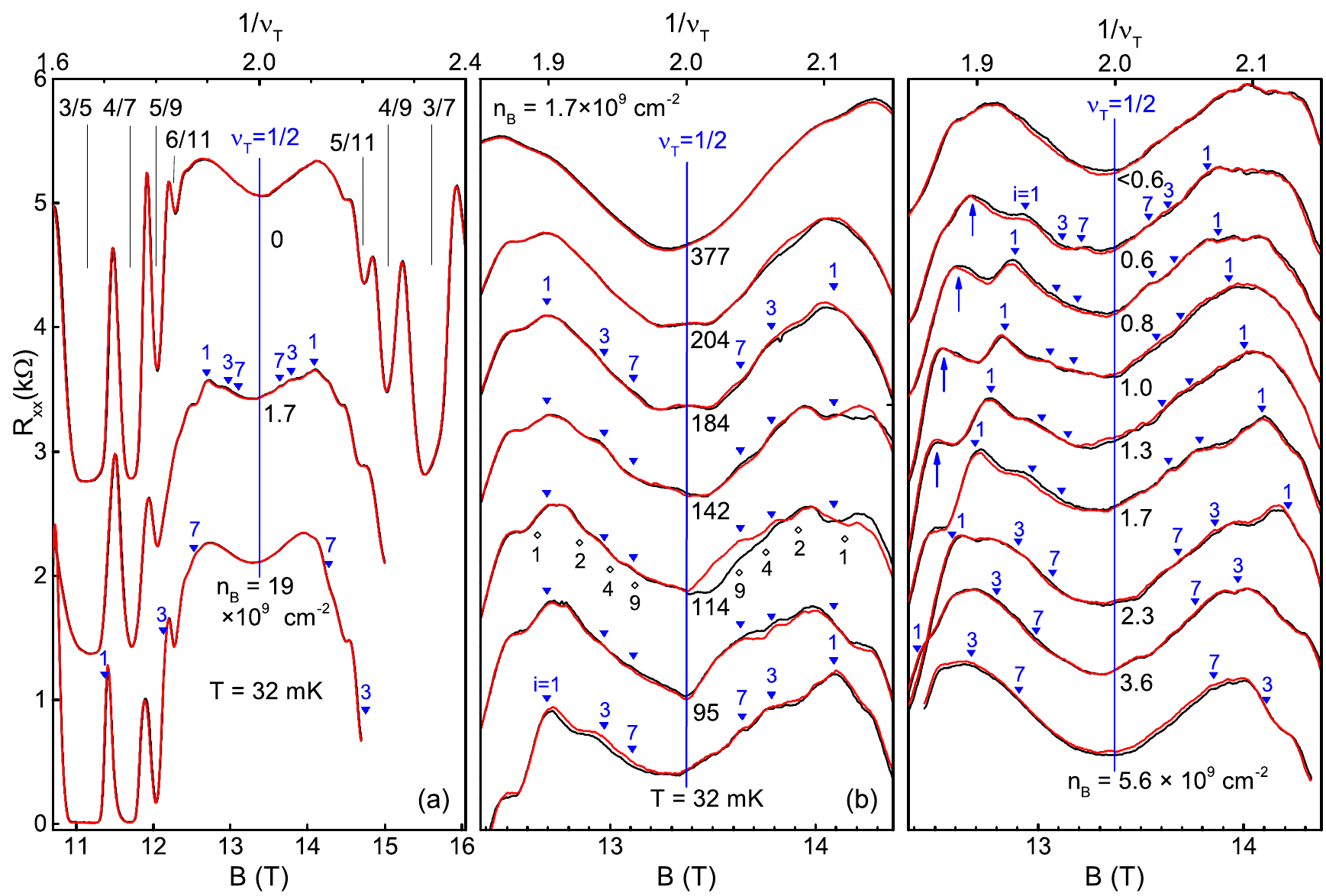}
\caption{
(a) $R_{xx}$ vs $1/\nu_T$ traces for the bilayer sample,
measured as the bottom-layer density is reduced via applying negative back-gate voltage.
The bottom-layer densities ($n_B$) are 19, 1.7, and zero (in units of 10$^{9}$ cm$^{-2}$) for $V_{BG}=0, -45$ and $-90$ V, respectively.
The top-layer density ($n_T$) for these traces varies slightly (see Fig. 3);
the top x-axis is normalized to represent $1/\nu_T$ where $\nu_T$ is the top-layer filling factor
(the lower axis (field) scale is only for the middle trace).
The triangles mark the \textit{expected} magnetic field positions
for the $i=1, 3, 7$ CF cyclotron orbit commensurability with a triangular WC potential (Fig. 1(b));
these positions are based on the \textit{measured} $n_T$ and $n_B$.
Note that for the lowest trace, the triangles fall far from $\nu_T=1/2$,
making the expected commensurability-induced resistance maxima indiscernible because of the strong FQHE features.
(b) Temperature dependence of $R_{xx}$ traces near $\nu_T=1/2$.
All traces were taken for $n_B = 1.7 \times 10^{9}$ cm$^{-2}$ ($V_{BG}=-45$ V).
For the $T=114$ mK trace, we also show the expected ($i=1, 2, 4, 9$) commensurability positions
if the WC had square symmetry.
(c) $R_{xx}$ vs $1/\nu_T$ traces are shown near $\nu_T=1/2$ for different $n_B$, corresponding to $-60\leq V_{BG}\leq-20$ V.
The expected positions of commensurability $R_{xx}$ maxima are indicated by triangles.
In all traces for $0.6\leq n_B\leq1.7 \times 10^{9}$ cm$^{-2}$,
note the presence of strong $R_{xx}$ maxima at the expected positions
for the $i=1$ commensurability for $\nu_T>1/2$.
Similar to (a), the lower axis (field) scale is only for the $n_B = 1.7 \times 10^{9}$ trace.
In (a), (b) and (c) two traces are shown for each condition:
the black trace is for up-sweep of \textit{B} and the red trace is for down-sweep.
The origin of the slight hysteresis seen in some of the traces is unknown.
Traces for different conditions are shifted vertically for clarity.
}
\end{figure*}

The longitudinal resistance ($R_{xx}$) vs $B$ traces presented in Fig. 2 capture the highlight of our findings.
Data are shown only in the high-$B$ region of interest.
In Fig. 2(a), we present traces for the bilayer sample at different $V_{BG}$.
In the top trace, the bottom-layer is completely depleted so that the trace represents $R_{xx}$ for the top-layer.
This trace exhibits the usual characteristics seen in high-quality 2D electron systems (2DESs):
$R_{xx}$ has minima at numerous odd-denominator fillings such as 3/5, 4/7, etc., indicating FQHE states,
and is essentially featureless close to $\nu_T=1/2$ where it shows a broad, smooth minimum.
The features in the lowest trace, are also characteristic of a simple, single-layer 2DES.
They, too, essentially reflect the properties of the top-layer,
including its density and filling factors as a function of $B$.
The bottom-layer, although present, is at a very small filling factor ($\nu_B \approx 0.05$),
has very high resistance,
and appears not to contribute to $R_{xx}$ at high fields near $\nu_T=1/2$.
(Based on measurements on single-layer samples, we esitmate the bottom-layer resistance to be several $100$ k$\Omega$ at high \textit{B}, about 100 times larger than the top-layer resistance.)
The middle trace in Fig. 2(a) which is taken just before the bottom-layer is depleted,
however, is unusual as near $\nu_T=1/2$ there are small oscillatory features
superimposed on the smooth background.
Moreover, as illustrated in Fig. 2(b), these features disappear when we warm up the sample.
As we discuss below, these oscillations near $\nu_T=1/2$
reflect the commensurability of the cyclotron orbits of the CFs in the top-layer
with the periodic potential induced by a WC formed in the bottom-layer (see Fig. 1).

To test our hypothesis, we carefully analyze the positions
of the anomalous $R_{xx}$ maxima observed near $\nu_T=1/2$
in the middle trace of Fig. 2(a);
this trace is shown enlarged near $\nu_T=1/2$ in Fig. 2(b) (lower trace).
A quantitative prediction of the expected positions of the CF commensurability features
induced by the WC potential requires accurate knowledge of both CF and WC layer densities, i.e.,
both top- and bottom-layer densities ($n_T$ and $n_B$) at high magnetic fields.
Fortunately, we can experimentally determine these densities quite precisely.
Our determination of $n_{T}$ is based on the magnetic field positions of the well-defined FQHE minima
observed near $\nu_T=1/2$.
For $n_B$, we subtract $n_T$ from the total density of the bilayer system ($n_{tot}$),
which we can also determine experimentally as described later in the manuscript.

Using $n_T$ for the density of CFs ($n_{CF}$) and $n_B$ for the density of WC electrons ($n_{WC}$),
we can predict the positions of commensurability features expected
when the CF cyclotron orbit circles around $i=1, 3, 7$ WC lattice points (Fig. 1(b)).
Assuming that each such orbit has equal distance to the nearest lattice points inside and outside the orbit (Fig. 1(b)),
the diameters for the $i=1, 3, 7$ orbits have values of $(1, \sqrt{3}, \sqrt{3}+1)$
in units of the period $a$ of the triangular lattice.
In the case of \textit{electrons} near zero magnetic field moving in a triangular anti-dot lattice,
strong magneto-resistance maxima consistent with such cyclotron diameter values have indeed been reported
\cite{Yamashiro.SSC.1991, Weiss.SurfSci.1994, Meckler.PRB.2005}.
The detailed nature and strength of the potential imposed by the WC on the CFs is of course unclear.
However, experimental data for both electrons \cite{Lorke.PRB.1990} and CFs \cite{Smet.PRB.1997, Smet.PRL.1998} indicate that,
even in the case of a \textit{weak} 2D periodic potential,
the $i=1$ CF commensurability resistance peak is observed when the cyclotron orbit diameter equals $a$.
In Fig. 2, we use small triangles to mark the positions of the \textit{expected} commensurability resistance maxima for the $i=1, 3, 7$ orbits.
These positions are based on the relation $R_{CF}=\hbar k_{CF}/eB_{eff}$,
where $R_{CF}$ is the CF cyclotron radius and $k_{CF}=\sqrt{4\pi n_{CF}}$ is the CF Fermi wavevector.
We use $n_{CF}=n_{T}$ for $B_{eff}>0$ and $n_{CF}=\frac{1-\nu}{\nu}n_T$ for $B_{eff}<0$,
namely we assume that $n_{CF}$ is equal to the minority carriers density in the lowest Landau level of the top layer; see Ref. \cite{Kamburov.PRL.2014}.
This leads to a slight ($\sim$ 5\%) asymmetry in the expected positions of the two CF resistance maxima for $B_{eff}>0$ and $B_{eff}<0$.

In Fig. 2(b), despite some clear discrepancies, there is overall a good agreement between the positions of the observed and expected resistance maxima.
We emphasize that the positions of the triangles are based on our \textit{measured} top- and bottom-layer densities and do not rely on any fitting parameters.
Note also the disappearance of the oscillatory features above $T\approx 200$ mK, which we associate with the melting of the WC.
We believe that the data in Fig. 2 provide evidence that the CFs in the top-layer are indeed dancing to the tune of a WC formed at low temperatures in the bottom-layer.

\begin{figure}[htbp]
\includegraphics[width = 0.4\textwidth]{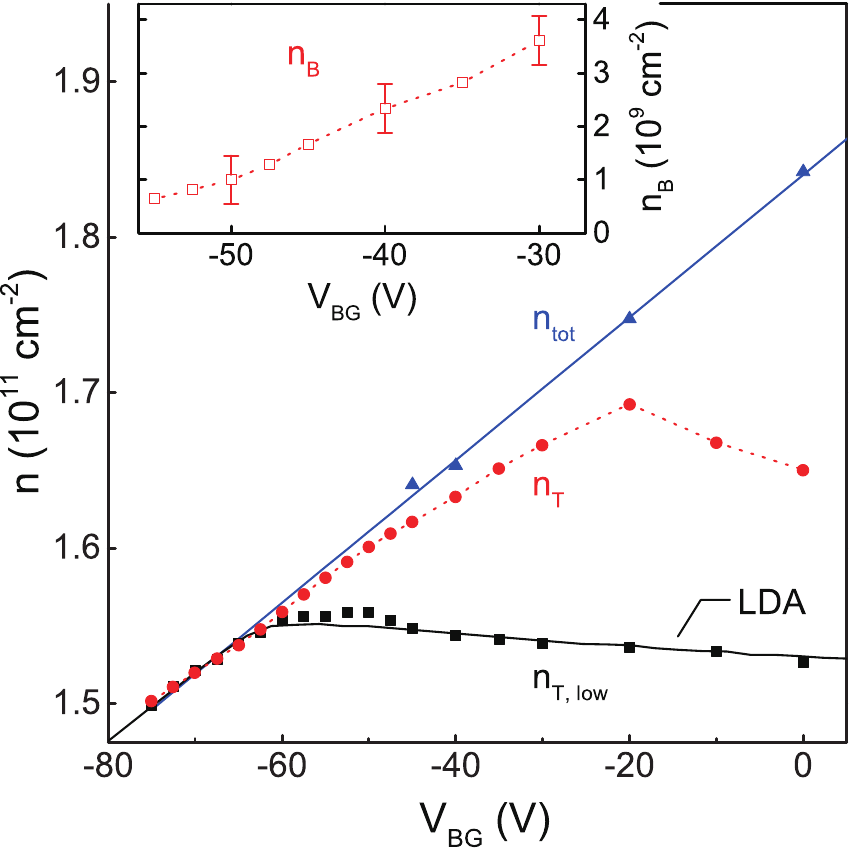}
\caption{
Measured densities for the bilayer system.
The blue triangles and black squares denote the total bilayer density ($n_{tot}$)
and the low-field top-layer density ($n_{T,low}$),
both determined from the Fourier transforms of the low-field Shubnikov-de Haas oscillations.
The solid black curve is the result of our self-consistent, local-density-approximation (LDA) calculations.
The red circles represent the high-field top-layer density ($n_T$)
determined from the field positions of the FQHE resistance minima near $\nu_T=1/2$.
In the inset, the red squares represent the bottom-layer density ($n_B$) at high fields,
deduced by subtracting $n_T$ from $n_{tot}$.
}
\end{figure}

We now describe details of our determination of top- and bottom-layer densities.
Besides providing $n_T$ and $n_B$, the data also reveal subtle interlayer charge transfers
that occur near zero and at intermediate \textit{B}.
In our experiments, for each value of $V_{BG}$,
we first measure the Shubnikov-de Haas oscillations at very low fields ($B\leq 0.7$ T).
The Fourier transform of these oscillations typically exhibits two peaks whose frequencies
directly yield the \textit{low-field} densities of the bottom and top layers ($n_{B,low}$ and $n_{T,low}$).
The sum of $n_{B,low}$ and $n_{T,low}$ gives $n_{tot}$, which we assume is independent of the magnetic field.
In Fig. 3, we show the measured $n_{tot}$ (triangles) and $n_{T,low}$ (squares) vs $V_{BG}$.
As expected, $n_{tot}$ changes linearly with $V_{BG}$
(with a slope that is consistent with the capacitance of our sample),
and $n_{T,low}$ equals $n_{tot}$ for $V_{BG} < -65$ V when the bottom-layer is completely depleted.
However, as $V_{BG}$ is decreased from zero,
$n_{T,low}$ \textit{increases} initially before decreasing at smaller $V_{BG}$.
This is because of the negative compressibility of an interacting 2DES,
and can be explained quantitatively using self-consistent (Poisson-Schroedinger equations) calculations
that include exchange and correlation energies via local-density-approximation
\cite{Eisenstein.PRB.1994, Ying.PRB.1995}.
In Fig. 3, we show the results of such calculations for the parameters of our bilayer electron sample,
indicating good agreement between the measured and calculated $n_{T,low}$ (black squares and curve in Fig. 3, respectively).

Surprisingly, however, in Fig. 3 we find that
the top-layer density at high magnetic fields ($n_T$)
is significantly larger than its low-field value $n_{T,low}$.
Our data imply that there is an \textit{additional} charge transfer
which is of the order of $\approx 10^{10}$ cm$^{-2}$ and occurs from the bottom- to the top-layer
at intermediate \textit{B}.
The origin of this charge transfer is the mismatch between the ground-state energies of the two layers
that have unequal densities but are in thermal equilibrium,
and the requirement that the lowest Landau levels of the two layers have to align at high magnetic fields
when both layers are in the extreme quantum limit.
In a simple, capacitive model, the magnitude of this charge transfer
is $\delta n = \Delta n\cdot \frac{\pi \hbar^2}{m^*}\cdot \frac{1}{\epsilon d}$,
where $\Delta n$ is the difference between the layer densities,
$d$ is the distance between the charge distribution peaks of the two layers,
and $\epsilon$ is the GaAs/AlAs dielectric constant.
While this simple model gives the correct order of magnitude for the observed charge transfer,
it does not quantitatively explain its size nor its dependence on $V_{BG}$.
The details of this subtle charge transfer are very interesting on their own
as they reflect a delicate balance between interaction at intermediate \textit{B}
and the capacitive energy associated with interlayer charge transfer.
We emphasize that this charge transfer has no bearing on our analysis of Fig. 2 data or our conclusions
because our determination of the CF and WC layer densities is based on direct measurements of $n_T$
(from the positions of FQHE minima near $\nu_T=1/2$)
and $n_{tot}$ (from Shubnikov-de Haas oscillations at low field),
and the reasonable assumption that $n_B = (n_{tot} - n_T)$.

Returning to the commensurability data,
in Fig. 2(c) we show additional traces taken near the depletion of the bottom-layer.
For each trace we indicate $n_B$ and also mark the expected positions of commensurability resistance maxima
based on $n_T$ and $n_B$ for that trace.
In a narrow range of $n_B$, oscillatory features are seen.
For these traces, the position of the expected $i=1$ maximum matches remarkably well the strong peak
observed to the left of $\nu_T=1/2$ ($\nu_T>1/2$).
It is noteworthy that experiments on commensurability of CFs in an anti-dot lattice
also show that the resistance maximum is more pronounced for $\nu>1/2$
compared to $\nu<1/2$ \cite{Kang.PRL.1993, Smet.PRB.1997}.
The reason for this asymmetry is, however, not known.
Similar to the data of Fig. 2(b), in some traces of Fig. 2(c) there are also hints of weak resistance maxima,
closer to $\nu_T=1/2$ compared to the $i=1$ commensurability maxima.
These could signal the commensurability of larger CF cyclotron orbits ($i=3$ and 7)
with the WC periodic potential.
In Fig. 2(c) traces, for $0.6\leq n_B\leq1.3 \times 10^{9}$ cm$^{-2}$,
there are also strong $R_{xx}$ maxima observed to the left of the $i=1$ triangles (see up-arrows in Fig. 2(c)).
We do not know the origin of these anomalous maxima;
they might indicate the resonance of a (small) CF cyclotron orbit that fits \textit{inside}
three adjacent lattice points.
We note that anomalous shoulders at fields highers than the $i=1$ resistance maximum are also seen in some commensurability data in anti-dot samples; see, e.g., traces 1 and 2* in Fig. 2 of Ref. \cite{PRL.66.2790.Weiss}.

So far we have assumed that the WC attains a triangular lattice.
This is indeed generally conjectured for a fully spin-polarized WC
\cite{Lozovik.JETP.1975, Lam.PRB.1984, Levesque.PRB.1984, Bonsall.PRB.1977}.
Near the 114-mK-trace in Fig. 2(b),
we have included marks (diamonds) for the expected positions of resistance maxima
if the WC lattice were square instead of triangular.
It appears that the positions of the triangles agree with the experimental data better than those of the squares.

Finally, the temperature dependent data of Fig. 2(b)
indicate the disappearance of the oscillatory resistance features above $T \approx 200$ mK,
signaling a melting of the WC.
A melting temperature of $\approx 200$ mK is comparable to but somewhat smaller than those
reported from various measurements \cite{Note3, Goldman.PRL.1990, Li.PRL.1991, Chen.NatPhys.2006}.
It is worth noting, however,
that the WC density in our sample ($n_{WC} \approx$ 1.7 $\times 10^{9}$ cm$^{-2}$ at $V_{BG}=-45$ V)
is about an order of magnitude smaller than the densities in previous measurements.
Also, it is possible that the CFs in the top-layer,
which is only $\approx 45$ nm away,
screen and reduce the Coulomb interaction between the electrons in the WC layer \cite{Ho.Hamilton.PRB.2009}.
Such screening would lead to a lower WC melting temperature.

Before closing, we remark on data we obtained from large (4 mm x 4 mm) van der Pauw samples.
Such samples exhibit more pronounced resistance maxima near $\nu_T=1/2$,
some of which can be assigned to the commensurability of top-layer CF cyclotron orbits with the potential
of a WC formed in the bottom-layer.
The WC layer density deduced from this assignment, however, is typically
much smaller (by a factor of up to ~4) than the bottom-layer density we obtain from our layer density measurements.
The discrepancy might stem from the non-uniformity of the layer densities in van der Pauw samples
whose area is about 100 times larger than the Hall bar samples presented here.
We note that density uniformity is especially important in our experiments
because the WC layer density has to be extremely small (of the order of $10^{9}$ cm$^{-2}$)
so that the CF commensurability condition is satisfied very near $\nu_T=1/2$, i.e., far from the FQHE minima.

Our results offer a direct look at the microscopic structure of the magnetic-field-induced WC in 2DESs.
They also raise several intriguing questions.
For example, what is the exact shape of the periodic potential modulation
that the WC layer imposes on the CFs?
A related question is the optimum interlayer distance that would lead to the most pronounced $R_{xx}$
commensurability maxima.
More generally, the weak periodic potential
imposed by the WC and the homogeneous \textit{effective} magnetic field $B_{eff}$
can conspire to also lead to a fractal energy diagram (the Hofstadter's butterfly)
\cite{Hofstadter.PRB.1976} for the of CFs,
similar to what has been studied for \textit{electrons} in an externally imposed periodic potential
\cite{Albrecht.PRL.2001, Melinte.PRL.2004}.
Finally, our technique might find use in studying other possibly ordered,
broken-symmetry states of 2DESs.
For example, we envision measurements where the CFs in one layer are used to probe,
in an adjacent layer,
the anisotropic phases that are observed at half-filled Landau levels with high index
\cite{annurev.conmatphys.070909.103925.Fradkin}.
These anisotropic phases are believed to signal many-body states,
consisting of stripes of electrons with alternating density (filling).
If the stripes are periodic,
they would generate a one-dimensional periodic potential in the nearby CF layer,
and the CFs should exhibit commensurability features.

\section*{Acknowledgements}

We acknowledge support through the NSF (DMR-1305691) for measurements,
and the Gordon and Betty Moore Foundation (Grant GBMF4420),
Keck Foundation,
the NSF MRSEC (DMR-1420541),
and the DOE BES (DE-FG02-00-ER45841) for sample fabrication.
We thank R.N. Bhatt and L.W. Engel for illuminating discussions.

\bibliographystyle{h-physrev}

\begin{thebibliography}{10}

\bibitem{Tsui.PRL.1982}
D.~C. Tsui, H.~L. Stormer, and A.~C. Gossard,
\newblock Phys. Rev. Lett. {\bf 48}, 1559 (1982).

\bibitem{Note1}
S.~D. Sarma and A.~Pinczuk, editors,
\newblock {\em Perspectives in Quantum Hall Effects} (Wiley, New York, 1997).

\bibitem{Note2}
M.~Shayegan,
\newblock Flatland Electrons in High Magnetic Fields,
\newblock in {\em High Magnetic Fields: Science and Technology, Vol. 3}, edited
  by F.~Herlach and N.~Miura,  (World Scientific, Singapore, 2006), pp. 31--60.

\bibitem{Jain.CF.2007}
J.~K. Jain,
\newblock {\em Composite Fermions} (Cambridge University Press, Cambridge, UK,
  2007).

\bibitem{Jain.PRL.1989}
J.~K. Jain,
\newblock Phys. Rev. Lett. {\bf 63}, 199 (1989).

\bibitem{Halperin.PRB.1993}
B.~I. Halperin, P.~A. Lee, and N.~Read,
\newblock Phys. Rev. B {\bf 47}, 7312 (1993).

\bibitem{Willett.PRL.1993}
R.~L. Willett, R.~R. Ruel, K.~W. West, and L.~N. Pfeiffer,
\newblock Phys. Rev. Lett. {\bf 71}, 3846 (1993).

\bibitem{Kang.PRL.1993}
W.~Kang, H.~L. Stormer, L.~N. Pfeiffer, K.~W. Baldwin, and K.~W. West,
\newblock Phys. Rev. Lett. {\bf 71}, 3850 (1993).

\bibitem{Smet.PRB.1997}
J.~H. Smet {\em et~al.},
\newblock Phys. Rev. B {\bf 56}, 3598 (1997).

\bibitem{Smet.PRL.1998}
J.~H. Smet, K.~von Klitzing, D.~Weiss, and W.~Wegscheider,
\newblock Phys. Rev. Lett. {\bf 80}, 4538 (1998).

\bibitem{Kamburov.PRL.2014}
D.~Kamburov {\em et~al.},
\newblock Phys. Rev. Lett. {\bf 113}, 196801 (2014).

\bibitem{Wigner.PR.1934}
E.~Wigner,
\newblock Phys. Rev. {\bf 46}, 1002 (1934).

\bibitem{Note3}
For a review, see M. Shayegan, "Case for the Magnetic-field-induced
  Two-dimensional Wigner Crystal", in Ref. \cite{Note1}; pp. 343-383.

\bibitem{Lozovik.JETP.1975}
Y.~E. {Lozovik} and V.~I. {Yudson},
\newblock JETP Lett. {\bf 22}, 11 (1975).

\bibitem{Lam.PRB.1984}
P.~K. Lam and S.~M. Girvin,
\newblock Phys. Rev. B {\bf 30}, 473 (1984).

\bibitem{Levesque.PRB.1984}
D.~Levesque, J.~J. Weis, and A.~H. MacDonald,
\newblock Phys. Rev. B {\bf 30}, 1056 (1984).

\bibitem{Andrei.PRL.1988}
E.~Y. Andrei {\em et~al.},
\newblock Phys. Rev. Lett. {\bf 60}, 2765 (1988).

\bibitem{Jiang.PRL.1990}
H.~W. Jiang {\em et~al.},
\newblock Phys. Rev. Lett. {\bf 65}, 633 (1990).

\bibitem{Goldman.PRL.1990}
V.~J. Goldman, M.~Santos, M.~Shayegan, and J.~E. Cunningham,
\newblock Phys. Rev. Lett. {\bf 65}, 2189 (1990).

\bibitem{Li.PRL.1991}
Y.~P. Li, T.~Sajoto, L.~W. Engel, D.~C. Tsui, and M.~Shayegan,
\newblock Phys. Rev. Lett. {\bf 67}, 1630 (1991).

\bibitem{Li.PRL.1997}
C.-C. Li, L.~W. Engel, D.~Shahar, D.~C. Tsui, and M.~Shayegan,
\newblock Phys. Rev. Lett. {\bf 79}, 1353 (1997).

\bibitem{Chen.NatPhys.2006}
Y.~P. Chen {\em et~al.},
\newblock Nature Phys. {\bf 2}, 452 (2006).

\bibitem{Tiemann.NatPhys.2014}
L.~Tiemann, T.~D. Rhone, N.~Shibata, and K.~Muraki,
\newblock Nature Phys. {\bf 10}, 648 (2014).

\bibitem{Yamashiro.SSC.1991}
T.~Yamashiro {\em et~al.},
\newblock Solid State Communications {\bf 79}, 885  (1991).

\bibitem{Weiss.SurfSci.1994}
D.~Weiss, K.~Richter, E.~Vasiliadou, and G.~Lutjering,
\newblock Surface Science {\bf 305}, 408  (1994).

\bibitem{Meckler.PRB.2005}
S.~Meckler {\em et~al.},
\newblock Phys. Rev. B {\bf 72}, 035319 (2005).

\bibitem{Lorke.PRB.1990}
A.~Lorke, J.~P. Kotthaus, and K.~Ploog,
\newblock Phys. Rev. B {\bf 44}, 3447 (1991).

\bibitem{Eisenstein.PRB.1994}
J.~P. Eisenstein, L.~N. Pfeiffer, and K.~W. West,
\newblock Phys. Rev. B {\bf 50}, 1760 (1994).

\bibitem{Ying.PRB.1995}
X.~Ying, S.~R. Parihar, H.~C. Manoharan, and M.~Shayegan,
\newblock Phys. Rev. B {\bf 52}, R11611 (1995).

\bibitem{PRL.66.2790.Weiss}
D.~Weiss {\em et~al.},
\newblock Phys. Rev. Lett. {\bf 66}, 2790 (1991).

\bibitem{Bonsall.PRB.1977}
L.~Bonsall and A.~A. Maradudin,
\newblock Phys. Rev. B {\bf 15}, 1959 (1977).

\bibitem{Ho.Hamilton.PRB.2009}
L.~H. Ho, A.~P. Micolich, A.~R. Hamilton, and O.~P. Sushkov,
\newblock Phys. Rev. B {\bf 80}, 155412 (2009).

\bibitem{Hofstadter.PRB.1976}
D.~R. Hofstadter,
\newblock Phys. Rev. B {\bf 14}, 2239 (1976).

\bibitem{Albrecht.PRL.2001}
C.~Albrecht {\em et~al.},
\newblock Phys. Rev. Lett. {\bf 86}, 147 (2001).

\bibitem{Melinte.PRL.2004}
S.~Melinte {\em et~al.},
\newblock Phys. Rev. Lett. {\bf 92}, 036802 (2004).

\bibitem{annurev.conmatphys.070909.103925.Fradkin}
E.~Fradkin, S.~A. Kivelson, M.~J. Lawler, J.~P. Eisenstein, and A.~P.
  Mackenzie,
\newblock Annual Review of Condensed Matter Physics {\bf 1}, 153 (2010).

\end{thebibliography}

\end{document}